%% file: proc.tex
\documentclass[a4paper,11pt]{article}
\usepackage{pos}
\usepackage[normalem]{ulem}

\usepackage{natbib}
\title{Results and performance of the Mini-EUSO telescope on board the ISS}
 \ShortTitle{The Mini-EUSO telescope on board the ISS}

\author*[a]{L. Marcelli}

\affiliation[a]{National Institute for Nuclear Physics, Division of Tor Vergata,\\
  Via della Ricerca Scientifica 1, I-00173, Rome, Italy}

\onbehalf{for the JEM-EUSO Collaboration\\[-1mm]{\normalsize \normalfont (a complete list of authors can be found at the end of the proceedings)}}
\emailAdd{laura.marcelli@roma2.infn.it}

\abstract{Mini-EUSO is a telescope observing the Earth in the ultraviolet band (290-430~nm) since 2019, through a nadir-facing UV-transparent window in the Russian Zvezda module of the International Space Station.
The main camera has an optical system composed of two 25~cm diameter Fresnel lenses and a focal surface consisting of 36 multi-anode photomultiplier tubes, 64~pixels each, for a total of 2304 channels. The instrument has a square field of view with a side of 44~degrees, a spatial resolution of about 6.3~km on the Earth surface and a sampling time of 2.5~microseconds. Mini-EUSO has also two cameras in the near infrared and visible ranges and silicon photomultiplier sensors to complement the UV observations. 
Mini-EUSO has been designed as a small-size version of the original JEM-EUSO space telescope to demonstrate its observational principle. Mini-EUSO is in fact potentially capable of observing extensive air showers generated by ultra-high-energy cosmic rays with an energy above 10$^{21}$~eV and of detecting artiﬁcial showers generated with lasers from the ground. 
Other main scientiﬁc objectives of the mission are the study of atmospheric phenomena (transient luminous events such as ELVES and sprites), the observation of meteors and among them the search  for interstellar meteors and nuclearites such as strange quark matter. Moreover, Mini-EUSO can map night-time UV Earth emissions, both anthropogenic and natural.
In this work, we will discuss results and performance of the telescope during its first four years of activity.}

\ConferenceLogo{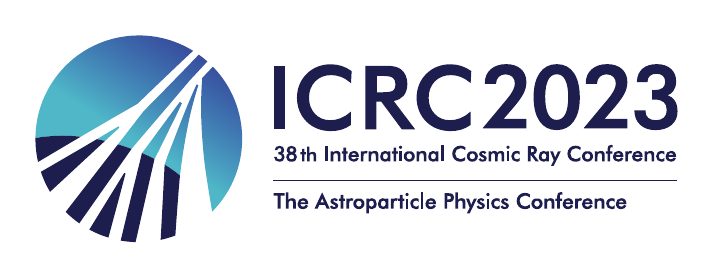}

\FullConference{%
38th International Cosmic Ray Conference (ICRC2023)\\
  26 July - 3 August, 2023\\
  Nagoya, Japan}

\begin{document}
\maketitle

\section{Introduction}

Mini-EUSO (Multiwavelength Imaging New Instrument for the Extreme Universe Space Observatory) experiment \cite{minieuso} is part of the JEM-EUSO program (Joint Exploratory Missions for an Extreme Universe Space Observatory) developed by the JEM-EUSO Collaboration \citep{Parizot:2023c4}. The aim of the Collaboration is to build a large space telescope to detect, for the first time, Ultra High Energy Cosmic Rays (UHECRs) from space. 

Over the years, the JEM-EUSO Collaboration has carried out many successful missions operating on ground (EUSO-TA \citep{2018APh...102...98A} (2013-)), on stratospheric balloons (EUSO-Balloon \citep{EUSO-BALLLOON-Adams2022,2019APh...111...54A} (2014), EUSO-SPB1 \citep{2017ICRC...35.1097W} (2017), EUSO-SPB2 \citep{Eser:2023Dw} (2023)), and in space (TUS \citep{Klimov2017} (2016)). Others are planned for the upcoming years: K-EUSO \citep{k-euso-universe} and POEMMA \citep{Olinto:2023ZQ}.

Mini-EUSO was launched to the International Space Station (ISS) onboard the unmanned Soyuz MS-14 on August 22, 2019, from the Baikonur Cosmodrome (Kazakhstan). The first installation of the instrument took place on October $7$, 2019, after the arrival on the ISS of the cosmonaut trained to operate the instrument (see Figure \ref{fig:fig1}, left-hand side). Since then, the telescope has been taking data periodically, with installations occurring every couple of weeks, for a total of about 90 sessions over four years. 

In this work, the performance of the instrument during these four years of data taking will be described together with its first results. 


\section{Instrument Description}

Mini-EUSO \citep{minieuso} has been designed to be installed in the interior of the ISS on the nadir-facing UV-transparent window located in the Russian Zvezda module. The dimensions (37x37x62 cm$^3$) are thus defined by the size of the window and the constraints of the  Soyuz spacecraft. Furthermore, the design accommodates the requirements of safety (no sharp edges, low surface temperature, robustness...) to the crew. Coupling to the window is done via a mechanical adapter flange; the only connection to the ISS is via 28 V power supply and grounding cables. The power consumption of the telescope is $\simeq$ 60 W and its weight is $\simeq$ 35 kg, including the  5 kg flange. The instrument has a square field of view with a side of 44~degrees, with a spatial resolution of about 6.3~km on the Earth surface.

For each observation session, taking place about every two weeks and of a duration of about 12 hours, the instrument is removed from the storage and installed on the UV-transparent window. Data are stored \citep{Data_acquisition_Software} on 512 GB USB Solid State Disks (SSDs) that are inserted on the side of the telescope by the cosmonauts. No direct  telecommunication with ground is present, but after each session a sample of the acquired data (about 10$\%$, usually corresponding to the beginning and the end of each session) is copied by the crew and transmitted  to ground via telemetry channel to verify the correct functioning of the instrument in order to optimize its working parameters. Conversely, before each session, updated working parameters can be uplinked to the ISS and then copied on the SSD disk to fine-tune the acquisition. Pouches with 25 SSDs are returned to Earth every 6-12 months and, with a similar time interval, another pouch with new SSD cards is sent to the station.

\begin{figure}[h]
\centering
\includegraphics[width=.99\textwidth]{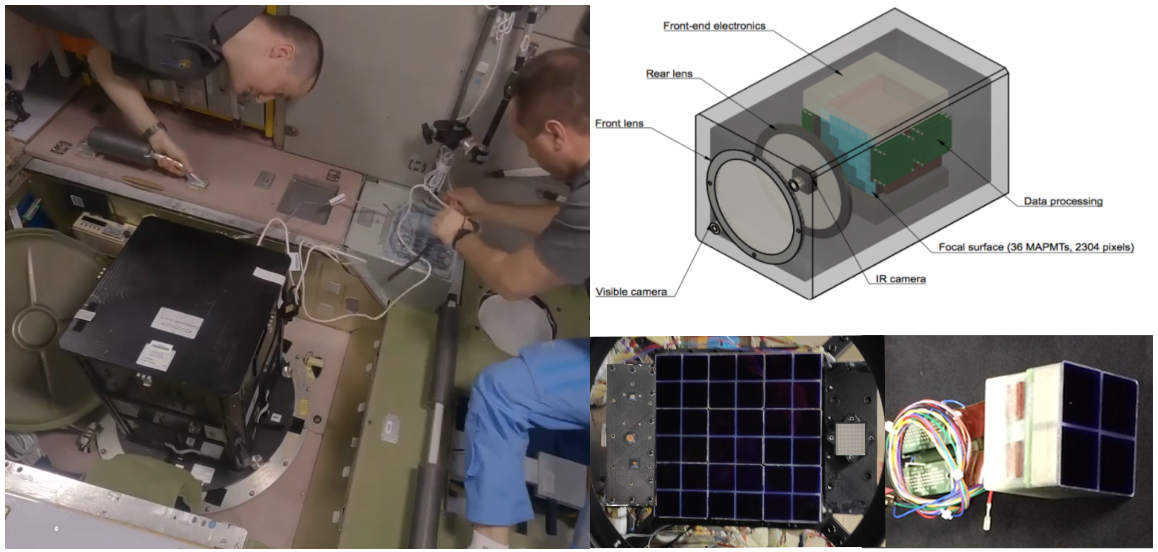} 
\caption{\textbf{Left:} Mini-EUSO installed inside the ISS on the nadir-facing UV-transparent window of the Zvezda
module by two cosmonauts. \textbf{Right-top:} schematic view of the instrument. \textbf{Right-bottom:} the focal plane composed of an 6x6 array of MAPMTs and a close-up of a single 2x2 module.}
\label{fig:fig1}
\end{figure}

The optical system is composed of two  Fresnel lenses with a  diameter of 25 cm. The Focal Surface (FS), or Photon Detector Module (PDM), consists of 36 MultiAnode Photomultipliers (MAPMTs) tubes by Hamamatsu, 64~pixels each, for a total of 2304 channels and single photon counting capabilities (Figure \ref{fig:fig1}, right-hand side). Readout is handled by ASICs (Application Specific Integrated Circuit) in frames of $2.5\: \mu s$ (this is defined as 1 Gate Time Unit, GTU). Data are then processed by a Zynq based FPGA board which implements a multi-level triggering \citep{Mini-EUSO_trigger}, allowing the measurement of triggered UV transients for 128 frames at time scales of both  $2.5\: \mu$s and $320\: \mu$s. An untriggered acquisition mode  with 40.96 ms frames performs continuous data taking. 
The instrument is also equipped  with two  ancillary cameras for complementary measurements in the near infrared and visible ranges and of a 8$\times $ 8 SiPM imaging array. Some UV sensors are used to manage the day-night transition during the data taking.

\section{Scientific Objectives}

As part of the JEM-EUSO program, Mini-EUSO has been developed to demonstrate the possibility of studying UHECRs from space. This means primarily proving that a space-based observatory has a sufficiently high duty cycle, intended as the fraction of time in which the atmospheric or anthropogenic light source do not obstacle the observation of a UHECR from space, as well as the capability to detect short light transients which show similarities in terms of either the light intensity or pulse duration with what is expected from an Extensive Air Shower (EAS) cascading in the atmosphere.
However, the size of the lenses (25 cm in diameter) implies a minimum energy threshold for UHECR detection well above $10^{21}$ eV (energy at which  EAS have so far not been observed).
Nevertheless, the collaboration plans to place an upper limit to the flux of particles at these energies using the fluorescence technique as Mini-EUSO has reached an exposure comparable to the one collected so far by ground-based experiments in hybrid mode~\citep{novotny,shin}.

\begin{figure}[h]
\centering
\includegraphics[width=0.6\textwidth]{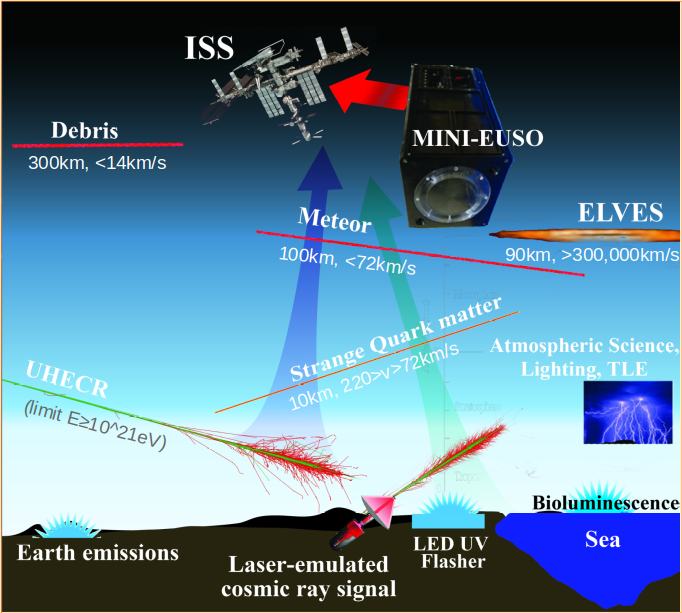}
\caption{Main scientific objectives of the Mini-EUSO experiment. The detector is capable of addressing a wide variety of phenomena with different durations and intensities, from the slow terrestrial emissions to the fast atmospheric events, such as ELVES.}
\label{fig-scientificobjectives}        
\end{figure}

Moreover, Mini-EUSO can contribute to search for exotic events that would not give a signal in the surface detectors, such as those observed by TUS~\citep{Khrenov2021}.
In addiction, measuring terrestrial and atmospheric UV emissions from the ISS orbit, Mini-EUSO allows us: to study atmospheric phenomena (including lightning and Transient Luminous Events (TLEs), such as ELVES); to observe meteors and meteoroids; to search for interstellar meteors and strange quark matter; to demonstrate the feasibility to detect and track space debris from space. Additionally, Mini-EUSO  acquires night-time terrestrial emissions in the ultraviolet range, both natural and anthropogenic. The main scientific objectives of Mini-EUSO are summarised in Figure \ref{fig-scientificobjectives}.

\begin{figure}[h]
\centering
\includegraphics[width=0.8\textwidth]{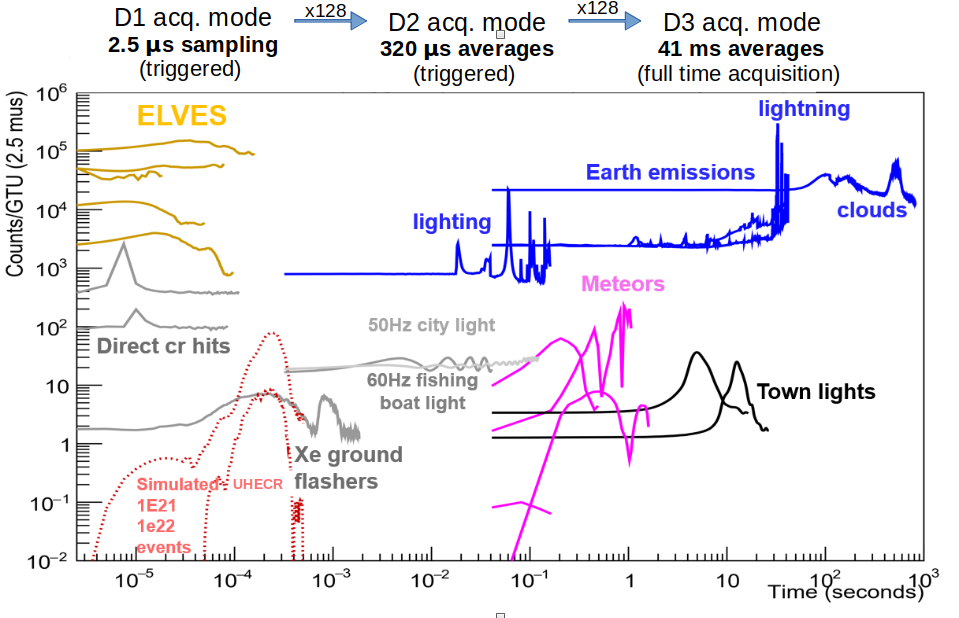}
\caption{\label{overview}Temporal profile of various signals observed by Mini-EUSO. All curves refer to experimental
data with the exception of the simulated UHECR events at 10$^{21}$ and 10$^{22}$ eV.}
\end{figure}

Figure \ref{overview} shows the total signal observed by the focal surface as a function of time for signals of various time scales, from the faster 2.5 $\mu$s sampling (D1 mode) to the 128 frame average for D2 acquisitions (320 $\mu$s) to the 128$\times$128 frame average for D3 mode (40.96 ms). 

The various signals detected by Mini-EUSO can be distinguished according to their temporal and spatial profile. In D1 acquisition mode we can distinguish fast events, such as direct hitting cosmic rays, ELVES, the blinking of the Xenon ground flasher, and we would be able to reveal UHECRs; in D2 time scale we observe the modulation of the artificial lights in small town and villages; and finally, using data acquired in D3 mode, we can study meteors, interstellar meteors and we can map the UV night-time Earth emissions, both natural and anthropogenic. 
 
\begin{figure}
    \centering
    \includegraphics[width=0.8\textwidth]{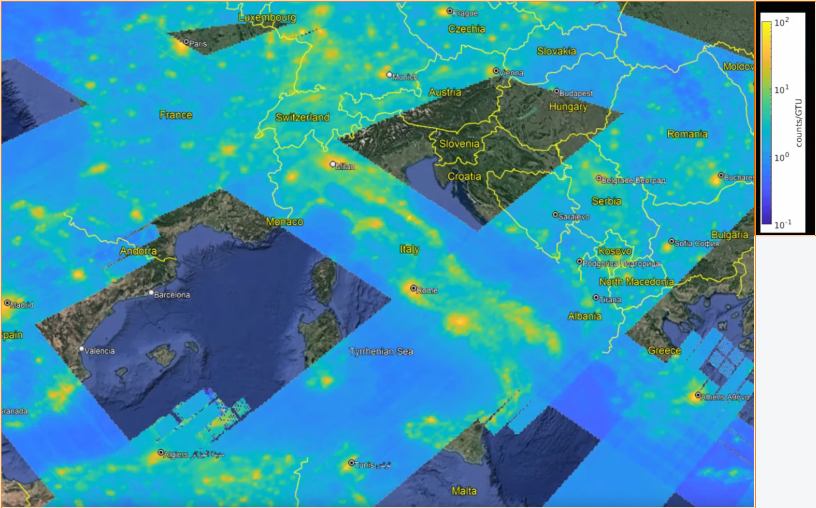}
    \caption{Map of Italy and part of Europe as reconstructed by Mini-EUSO. The coasts are well reconstructed and the major cities are clearly recognisable. On the ocean and on the land without urbanization the UV intensity is typically < 1 count/pixel/GTU).} 
    \label{fig:ItalyMap}
\end{figure}

The main types of  events, from the fastest to the slowest, are:

\begin{itemize}
\item \textbf{Direct hits} on the focal surface are due to cosmic rays that directly interact  with the photocathode or the BG3 filter either with direct ionization or emitting Cherenkov light. Most of these events cross one or a few pixels, releasing a high signal that lasts a few GTUs and exhibit a sharp increase and an exponential decrease due to the de-excitation of the elements hit. In Section \ref{sec:UHECRs} their signals are compared to that of UHECRs.
Figure \ref{Fig:DCR_and_EAS_like}, left panel, shows a direct hit with its typical exponential decrease.

\item \textbf{UHECRs}: as expected due to the extreme energy threshold combined with a short exposure, the search for UHECRs has so far yielded no results. However, the detection of short light transients demonstrates indirectly that the JEM-EUSO technology can detect UHECRs from space as they show similarities in terms of light profile, intensity, duration, and pixel pattern on the focal surface, even though all these characteristics do not match at the same time for a single event. More importantly, Mini-EUSO showed that those events can not be mistaken for real EAS-induced signals, and, therefore, do not represent a problem for future observations.
A large space is dedicated to this topic in Section \ref{sec:UHECRs}, as it is one of the topics of greatest interest to this Community.

\item \textbf{Ground flashers} are used as obstruction lights (usually with Xenon) to warn aircraft of the presence of buildings or towers. They have different brightness and blinking duration but usually last a few hundred $\mu s$ (see Figure \ref{overview}) and are usually observed by Mini-EUSO several times as they move in the field of view of the instrument \cite{ADAMS20141506}. 

\item \textbf{ELVES} are observed as fast-expanding luminous rings in the ionosphere lasting about half a millisecond. They belong to the family of the TLEs.
Mini-EUSO is able to observe the temporal and spatial development of the ring with a high resolution, allowing us to study ELVES in a very detailed way. Section \ref{sec:elves} is dedicated to this item.

\item \textbf{Light modulation} at 50 or 60 Hz of the artificial lights can be identified in D2 time scale. This modulation is  better visible in small towns and villages, which are all connected to the same transformer, than in larger cities, which have different sections connected to different transformers and with varying phases. Figure \ref{overview} shows the light modulation from India and Canada and from some fishing boats detected in the Indian Ocean. 

\item \textbf{Lightning} are atmospheric transient events with duration of $\simeq$ 1 s that can illuminate partially or entirely the focal surface, often triggering the high voltage safety system \citep{CASOLINO2023113336}. Passages over areas with high lighting activities can take some hundreds of seconds.
Figure \ref{overview} shows the time profile of some lightning events as seen in D2 and D3 modes.

\item \textbf{Meteors} are identified in the D3 time scale looking for straight tracks moving in the field of view. Meteors have signals varying in intensity and duration depending on mass, velocity and angle of incidence. Studying meteors, we also search for interstellar meteors and nuclearites. See Section \ref{sec:Meteorsand} for a more detailed discussion.

\item \textbf{Night-time UV Earth emissions} move in the field of view, and consequently on the focal surface, with an apparent speed close to the orbital velocity of the ISS ($\simeq$7.7 km/s). A given point on the Earth surface is thus visible for about 50 s ($\simeq 1000$ frames in D3 mode): it is thus possible to derive ground maps with good spatial resolution and reduced statistical fluctuations. 
The temporal behaviour of the signal on a given pixel depends on the size of the town and its neighbourhood. The typical time  profile of a single pixel (see Figure \ref{overview}) consists of a gradual growth of light lasting for some seconds, according to the size of the town.
Figure \ref{fig:ItalyMap} shows the map of Italy and part of Europe as reconstructed by Mini-EUSO. 
For an overview of Mini-EUSO capabilities in UV map reconstruction refer to \citep{CASOLINO2023113336}.
\end{itemize}  

\section{Some Mini-EUSO results}

\subsection{Duty cycle, exposure, UHECRs and Short Light Transients} \label{sec:UHECRs}

\begin{figure}[h]
\centering
\includegraphics[width=\columnwidth]{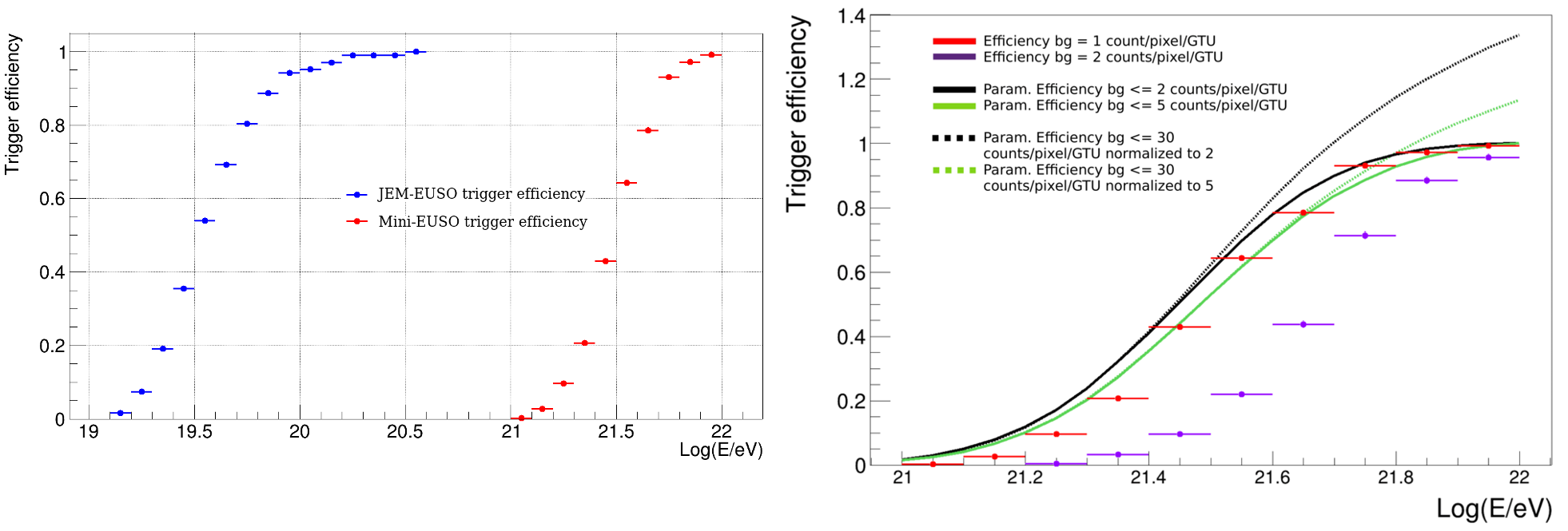}
\caption{Left: Comparison between the trigger efficiency curves of JEM-EUSO and Mini-EUSO using ESAF simulated proton-generated EASs of different energies. The background used for both experiments is the nominal one: 1.1 counts/pixel/GTU for JEM-EUSO and
1 count/pixel/GTU for Mini-EUSO. As expected the two curves are shifted by $\sim$2 orders of magnitude. 
Right: Trigger efficiency curves of Mini-EUSO for ESAF simulated proton-generated EASs of different
energies on different background levels.
The red and violet points assume a fixed nightglow background of 1 and 2 counts/pixel/GTU, respectively.
The continuous black (green) line represents the convolved trigger efficiency curve in which each background level below 2(5)
counts/pixel/GTU is weighted for the relative fraction of time in which it was measured by Mini-EUSO.
The dotted black (green) line provides the fractional increase in exposure (relative to the black (green) line) if
the accepted nightglow background is increased from 2(5) counts/pixel/GTU to 30 counts/pixel/GTU. 
A significant increase in exposure is obtained only at the highest energies. See text for details.
}
\label{fig:comparison-je-me}       
\end{figure}

In order to demonstrate the feasibility of detecting UHECRs from space, Mini-EUSO was designed to measure a photon rate per pixel from diffuse sources (nightglow, clouds, cities, etc.) in the range of values expected from a large mission in space, such as the original JEM-EUSO mission~\citep{exposure}. The pixel FoV is, therefore, $\sim$100 times larger in area with respect to the FoV of a JEM-EUSO pixel ($\sim$0.5 $\times$ 0.5 km$^2$), to compensate for the optical system $\sim$100 times smaller, constrained by the dimension of the UV transparent window. This implies that in case of point-like sources, the energy threshold of Mini-EUSO is $\sim$2 orders of magnitude higher.
A full simulation of JEM-EUSO and Mini-EUSO detectors was performed with ESAF~\citep{esaf2, esaf} to confirm a similar count rate from diffuse sources. In case of Mini-EUSO the overall efficiency of the detector was fine-tuned to match the measured one $\epsilon_{ME}$ = 0.080 $\pm$ 0.015 (see~\citep{Miyamoto:202312}). The result confirmed expectations with only a $\sim$5\% higher count-rate in JEM-EUSO.
Left side of Figure ~\ref{fig:comparison-je-me} shows the comparison between the efficiency curves of JEM-EUSO and Mini-EUSO for nominal background levels of 1.1 and 1.0 counts/pixel/GTU, respectively. As expected, at 50\% trigger efficiency there is a scaling factor of $\sim$100 in energy threshold betweeen JEM-EUSO and Mini-EUSO.

In~\citep{exposure} it was derived that effective duty cycle of JEM-EUSO mission, meant as the fraction of time in which the UV background light intensity (from either nightglow, moon phase, anthropogenic lights, lightning and aurorae) allows the UHECR observation, is $\sim$18\% and that 1.1 counts/pixel/GTU corresponds to the typical measured UV intensity.
The night-time Earth observations by Mini-EUSO are consistent with this assumption. In fact, in no-moon conditions, in more than 90\% of the time of clear sea conditions the count rate belongs to the interval 0.4 - 2.0 counts/pixel/GTU, the median being $\sim$0.8 counts/pixel/GTU
for both clear sea and land conditions. In clear land conditions there is a higher probability ($\sim$16\%) of very low counts (< 0.5 counts/GTU) in comparison with areas covering bodies
of water ($\sim$9\%). These areas of very low brightness are mainly distributed over forests and deserts (see~\citep{CASOLINO2023113336} for details). This indicates that these are areas where the background is at least twice lower compared to the nominal one allowing the detection of EASs of energies approximately $\sqrt{2}$ times lower than nominal ones, thus increasing the range of energies in which a space based observatory could overlap with ground-based ones in order to cross-check possible systematic dependence of the energy scale. 

Cloudy conditions typically shift curves by a factor 1.5 - 2 towards higher count rates as already measured in JEM-EUSO balloon flights~\citep{EUSO-BALLLOON-Adams2022}. Moreover, the chance probability to be in a cloudy region when the UV intensity is lower than 0.7 counts/pixel/GTU is less than 5\%. This means that $\sim$40\% of the clear sky conditions is already automatically defined by the fact of measuring a count rate < 0.7 counts/pixel/GTU. This result is very important because it allows performing consistency checks between the energy spectrum and anisotropy maps measured in these low background conditions with those adopting a larger data sample.

A procedure was defined in Mini-EUSO analysis to calculate a convolved triggering efficiency curve which takes into account the fraction of time in which each background level is measured (see~\citep{Bertaina:2023ok}). 
Right side of Figure \ref{fig:comparison-je-me} shows the convolved trigger efficiency curves
obtained by integrating all background levels below 2(5) counts/pixel/GTU.
 The result is shown as black(green) line. The green curve is very close to the efficiency curve obtained at
the fixed background level of 1 count/pixel/GTU in Mini-EUSO, indicating that such a curve provides a reasonable
approximation of the integrated performance in the various UV background conditions. The limit value of 5 counts/pixel/GTU represents the value at which the integrated duty cycle corresponds to 18\% as derived in JEM-EUSO estimations. Therefore, the present results demonstrate that a space-based observatory indeed has an effective duty cycle compatible with 18\%. This estimation doesn't include the reduction of the geometrical factor due to clouds which amounts to $k_c$ = 72\% as derived at the time of JEM-EUSO assuming the climatological distribution of clouds for the ISS orbit, which brings the effective conversion factor from aperture to exposure to $\eta$= 13\%. 

As Mini-EUSO observes the Earth only in limited sessions which are often chosen to maximise the duty cycle, the 18\% has been obtained by properly weighing the different measurements with the monthly moon phase occurrence. Without such weighting factor, the effective Mini-EUSO duty cycle increases to $\sim$25\%. The estimated exposure accumulated by Mini-EUSO during sessions 20 - 44, adopted in this analysis, corresponds to $\sim$1400 Linsley (including $k_c$). The projected accumulated exposure till the last session of Mini-EUSO (number 90 as of August 15$^{th}$ 2023) corresponds to $\sim$4000 Linsley. These values are comparable to the ones collected so far by ground-based experiments in hybrid mode~\citep{novotny,shin}. More details are reported in~\citep{Bertaina:2023ok}. 

\begin{figure}[h]
\centering
\includegraphics[width=.85\textwidth]{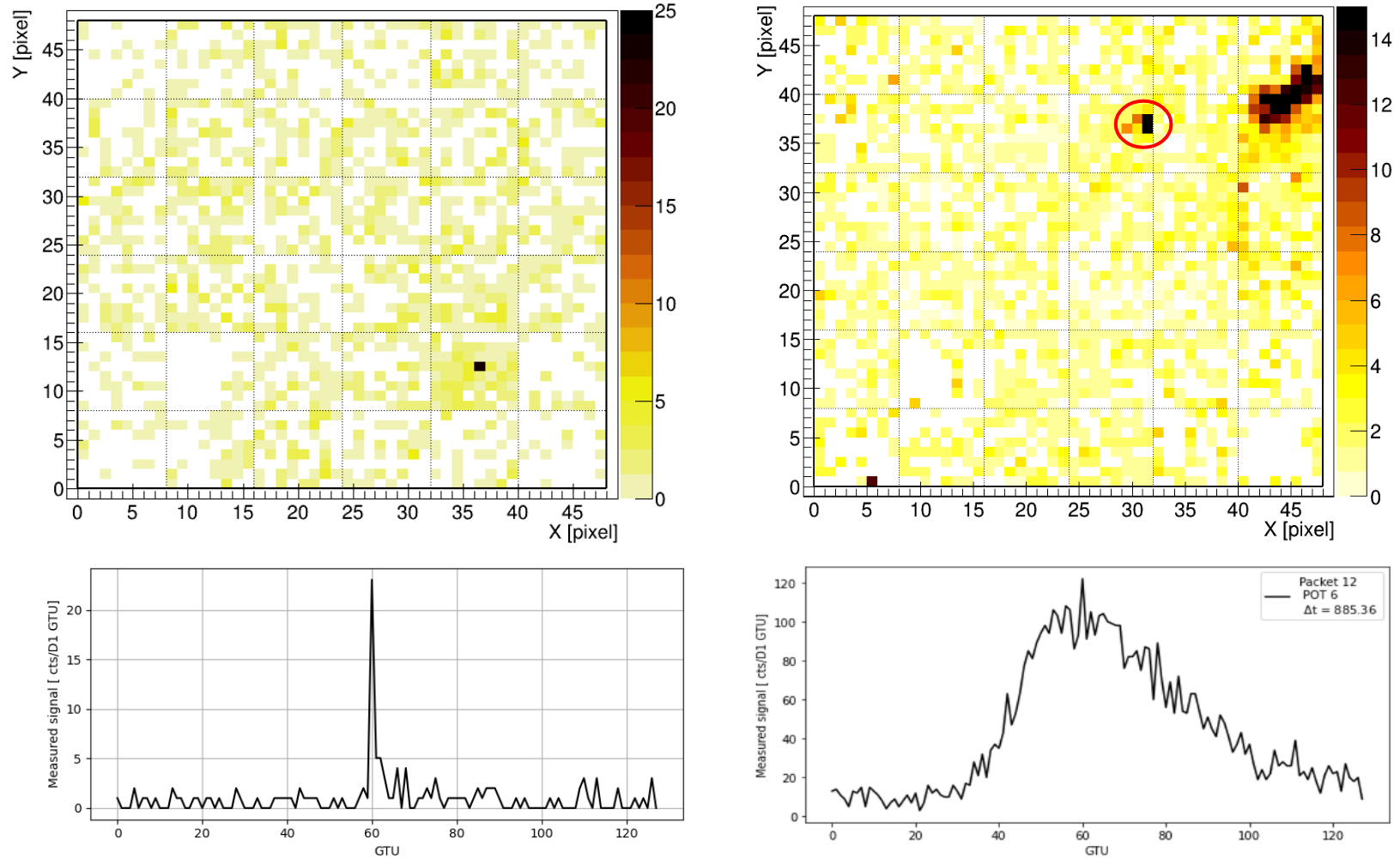}
\caption{\textbf{Left:} Example of a direct cosmic ray. The trigger is caused by a low energy ($\simeq$ GeV) cosmic ray impinging orthogonally to the focal surface of the detector and leaving a bright signal in one pixel (number of photoelectron counts/GTU are reported on the Z-axes). The corresponding bottom plot shows the light-curve of the brightest pixel. \textbf{Right:} Example of a EAS-like event (or, more properly, a Short Light Transient, SLT). The event has been detected off the coast of Sri Lanka (the bright area on the right of the focal plane). It appears in a small cluster of pixels and shows a bi-gaussian light-curve, with a faster rise and a slower decay. The light-curves shown in this work are the sum of all the pixels over threshold (POT in the legend) in the packet (6 in this case). }
\label{Fig:DCR_and_EAS_like}
\end{figure}

The trigger studies reported the expected functioning of the trigger logic with a trigger rate on spurious events within the requirements in nominal background conditions. The trigger logic proved effective avoiding excessive trigger rates in the presence of static anthropogenic lights such
as cities, while it increases significantly in presence of thunderstorms, and in areas of low geomagnetic cutoff where more direct cosmic rays are detected (see left side of Figure ~\ref{Fig:DCR_and_EAS_like}). Both cases do not pose serious issues for future large experiments. Thunderstorms are associated with the presence of high clouds when the UHECR observation is not feasible, while the trigger rate of direct hits will be significantly suppressed by the presence of a double level trigger rate in large experiments. More details are reported in~\citep{matteo-asr}.

\begin{figure}[hbtp]
\centering
\includegraphics[width=.98\textwidth]{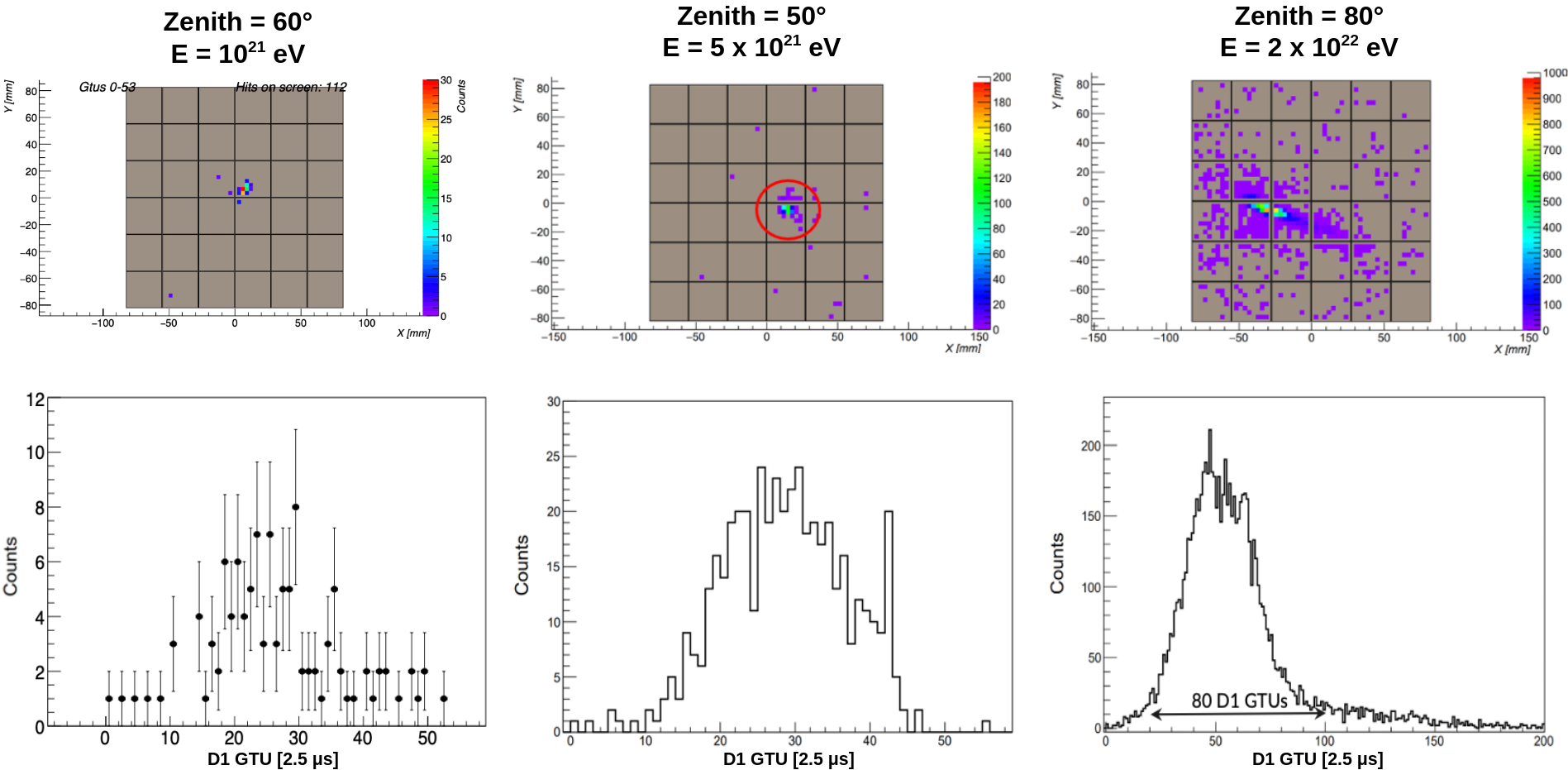} 
\caption{ESAF~\cite{esaf2, esaf} simulation of protons with different energies and zenithal angles. Given its high energy threshold, a shower of 10$^{21}$~eV \textbf{(Left)} is at the edge of the Mini-EUSO triggering capabilities. Around $5\times10^{21}$~eV the signal of a 50$^\circ$ shower (\textbf{Middle}) is clearly visible and lasts for $\sim$30~GTUs ($\sim$75~$\mu$s). The signal is truncated when the shower reaches the ground. At 80$^\circ$ zenith angle (\textbf{Right}, energy $2\times10^{22}$~eV) the light-curve is not truncated and the signal lasts for $\sim$80~GTUs ($\sim$200~$\mu$s). }
\label{fig:ESAF_simulation_Mini_EUSO}
\end{figure}

A search for EAS-like events was performed in Mini-EUSO data~\citep{Battisti:2023kN}. An EAS would appear as a signal persisting in a pixel for at least $\sim$8 GTUs and eventually moving to neighboring pixels, with a light profile that matches a bi-gaussian shape, with a faster rising and a slower decay that can be abruptly truncated for vertical events when the EAS reaches the ground. Three examples of EAS simulations are shown in Figure \ref{fig:ESAF_simulation_Mini_EUSO} for different energies and zenithal angles,
while Table~\ref{tab:L1trig} summarizes the main categories of events detected by the L1 trigger and their specific signatures.

The most interesting categories in the search for EAS-like events are Ground Flashers (GFs) and Short Light Transients (SLTs). They are both identified by searching for short events which can be repeated after few seconds or not. Sometimes they trigger two consecutive packets and the full shape is visible. Those with a duration $\leq200~\mu$s were looked more carefully to check their compatibility with EAS-like events.
The GF events are triggered several times (at least 3, but usually up to 15 or 20 times) in the $\sim$50~s during which a point on the ground stays inside the Mini-EUSO FoV. The signal is usually confined in a small cluster of pixels, while the intensity and duration can change a lot from case to case, and can, sometimes, resemble the signal expected by an UHECR. Their anthropogenic origin is, however, confirmed whenever the same event is triggered several times and they are never misidentified for EAS-induced signals, despite their large number.  Most likely, GF signals are produced by blinking lights on ground, usually located near airports, ports or cities.
Some of these events offer the opportunity to prove the sensitivity of space-based observatories to UHECRs in the energy range they are expected to be detected as these events show similarities in terms of either the time duration or the light shape and intensity with the expected signals from UHECRs, even though the combination of light excess on the FS and signal duration do not match at the same time the ones expected from EAS tracks.
It is important to underline that in a space-based observatory they would be more clearly identified as point-like sources because they brighten only on one or a few pixels while an EAS track of an almost vertical EAS extends on more than one MAPMT.

\begin{table}
\centering
\small
\caption{Categories of events in D1 data.}
\begin{tabular}{lll}
\hline
\hline
Category & Duration & Signature \\
\hline
\hline
Ground Flasher (GF) & Tens of D1 GTUs. & Bright spot suddenly appearing \\
 & & in one or few neighboring pixels,\\
 & & triggered many times as it moves\\
 & & in the FoV.\\
\hline
Short Light Transient & Between 30 and 80 D1 GTUs, & Bi-gaussian light-curve. Does not\\
(SLT) & with a faster rising and & present an apparent movement.\\
& slower decay. & Cosmic origin excluded through\\
& & comparison with simulations.\\
 \hline 
Direct cosmic ray & Rise time of 1 or 2 D1 GTUs. & Very different shapes, but\\
& Immediate or exponential decay& characterised by a peculiar \\
& shorter than 15 D1 GTUs.& light-curve. \\
\hline
ELVES & Usually 64 D1 GTUs, from trigger & Expanding ring-shaped event.\\
&to the end of the D1 packet. It can&\\
& be longer if it triggers more packets.&\\
\hline
\hline
\label{tab:L1trig}
\end{tabular}
\end{table}

The last category of interesting events are the SLTs which are represented by any flashing signal lasting no more than 200~$\mu$s that are not originated from a GF. In the currently analysed dataset 14 SLT candidates were identified. An example of their light curves is shown in the right side of Figure ~\ref{Fig:DCR_and_EAS_like}. Usually, the light-curves present a bi-gaussian shape, with a relatively long signal not fully contained in one packet. None of those 14 events presents an apparent movement on the focal plane but appears as a stationary light confined in a small cluster of pixels, switching on and reaching the maximum before starting to fade away. 
The origin of those fast flashing lights is still under study, but it seems safe to assume that at least some of them are linked with the thunderstorm activity in the atmosphere. For 6 SLT events, in fact Mini-EUSO detected an atmospheric event in the exact same pixels shortly after the SLT (between 1~ms and 200~ms). We are currently working on the identification of those 6 atmospheric events, that appear to belong to the class of TLEs
rather than to be more common thunder strikes. We are also investigating any possible correlation between these 6 or other Mini-EUSO atmospheric events with Terrestrial Gamma-ray Flashes (TGFs), which are known to be linked to thunderstorm activities.
The event was compared to different simulated EASs with variable energy and zenith angle. No simulated
EAS (in Figure \ref{fig:ESAF_simulation_Mini_EUSO}) is compatible with both the image size and the time duration of the light profile shown in Figure \ref{Fig:DCR_and_EAS_like}, right side. In fact, the light spot is compatible with a nearly vertical event, but the duration is much longer than the time needed by a vertical EAS to develop in atmosphere and reach the ground. This event has,
therefore, a different nature which is currently under investigation.
The D3 data of Mini-EUSO
indicate that the event occurred in a region between cloudy and clear atmospheric
conditions as the D3 video data show the passage of patches of clouds in that area. This observation is supported by Himawari satellite \citep{himawari} images and GFS \citep{GFS_cisl_rda_ds084.6} analysis and underlines the importance of D3 data to monitor the actual weather conditions when an interesting event occurs and in case raise an alert on the quality of the measured signal.

Finally, we searched in Mini-EUSO data for EAS-like events such as those detected by TUS and reported
in~\cite{Khrenov2021}. We found that GF signals might generate light
profiles similar to what was detected by TUS. 
However, we underline that in the TUS161003 event the
signal appears to be moving among pixels with a relativistic speed, while in Mini-EUSO data
all the signals with a duration compatible with EAS appear stationary (either induced by flashers or not). Thus, Mini-EUSO can not confirm yet the type of events detected by TUS. 
It is important, anyway, to underline that this search of EAS-like events has been performed on an estimated exposure of $\sim$1400 Linsley which is comparable to the one collected by TUS in EAS-mode 
(see~\cite{mario-tus}) and that TUS observed a few other EAS-like events even though not with the same quality as TUS161003. Therefore, under the assumption that the TUS event does not have an anthropogenic origin or that the movement of the track is artificially induced by some optical effect, one possible explanation could be related to the fact that the EAS-like events detected by TUS are estimated to have energies around 10$^{21}$ eV which is at the limit of the detection capability of Mini-EUSO.

\subsection{ELVES} \label{sec:elves}
\begin{figure}[!h]
    \centering
\includegraphics[width=0.85\textwidth]{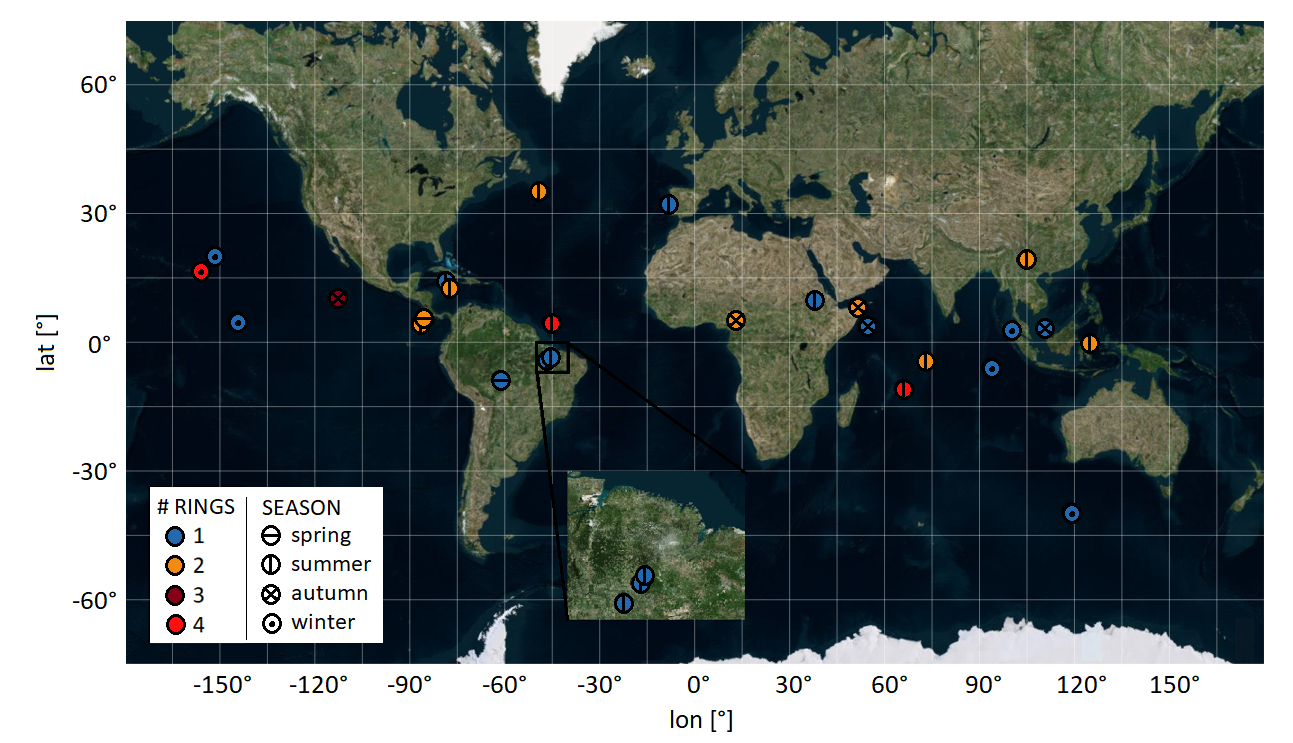}
    \caption{Location (latitude-longitude) of the ELVES detected with Mini-EUSO. Most of them are distributed in the equatorial region. Three events in East Brazil occurred at an interval of 18 and 5 seconds (second from first and third from second respectively) one from the other.}
    \label{fig:Location}
\end{figure}

Analysing the available dataset (half of the dataset is still on the ISS) we found that Mini-EUSO observed 37 ELVES. Figure \ref{fig:Location} shows the locations of the events that have been detected. As expected, most of them are located in regions close to the equatorial zone. About 50$\%$ of the detected events are single-ringed ELVES, 33$\%$ are double-ringed ELVES and the remaining part have three or more rings.

ELVES are detected by the Mini-EUSO focal surface as bright circle arcs of light rapidly expanding over time in the ionosphere \citep{Piotrowski:2023pN}. The typical ELVES lifetime is about 0.5 ms, this means that several $2.5 \;\mu \text{s}$ frames (on average 200) are associated with each event.
The temporal (2.5 \;$\mu s$) and spatial (5 km at the height of the ionosphere) resolutions of the instrument allow for a detailed study of the morphological structure of ELVES, enhanced by the fact that the telescope observes the expanding rings of light from a nadir point of view.

\begin{figure}[!h]
    \centering
    \includegraphics[width=0.90\textwidth]{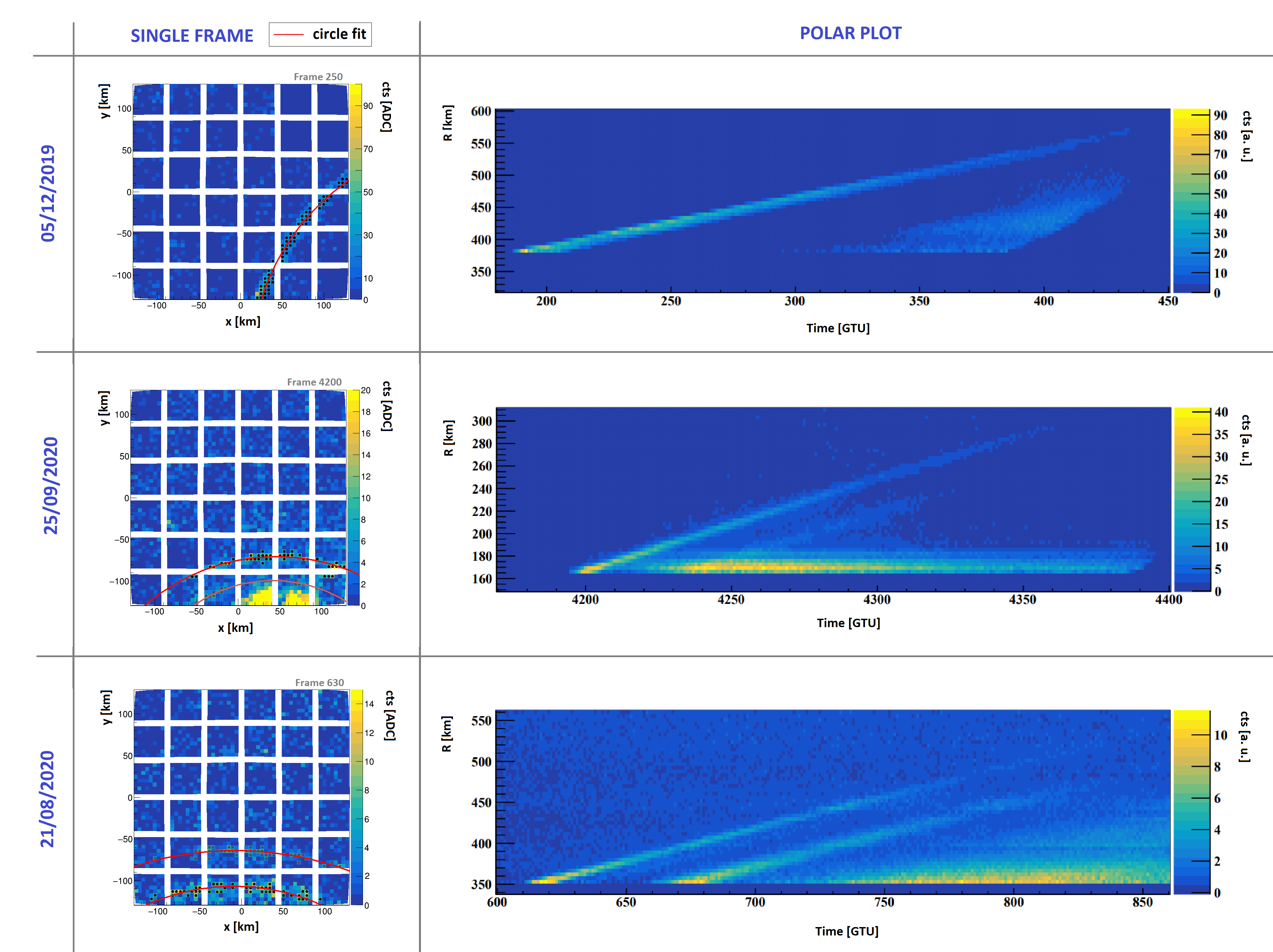}
    \caption{The signal detected in a single frame (2.5\; $\mu$s) on the focal surface (left column) and radius versus time, R(t), distribution (right column) for three ELVES observed by Mini-EUSO. In the R(t) distributions, ELVES appear as high-count inclined bands. From top to bottom: a) a single-ringed event; b) a double-ringed event, with a halo concurrent with the ELVES, static in time; c) a double-ringed event, with a halo following the ELVES and expanding at the same velocity.}
    \label{fig:ICRC2}
\end{figure}

Figure \ref{fig:ICRC2} shows, in the left column, the signal released in a single frame on the focal surface for three different events reported in the three rows: a single-ringed ELVES (first event) and two double-ringed ELVES (second and third events). The rings of the signals are fitted with red circles used to calculate the ELVES centers.
Once the position of the ELVES center is computed, it is possible to calculate the total number of photoelectron counts detected at a given time t as a function of the radial distance $R$ from the center. 
These distributions are shown in Figure \ref{fig:ICRC2}, right column. With this representation, ELVES are visualized as high-count inclined bands. Double-ringed ELVES are seen as parallel bands, and halos are sometimes visible after the ELVES as a delayed brightness (second and third events). The velocity of propagation of the rings can be estimated by looking at the corresponding $R(t)$ distribution and it is compatible with the speed of light. More
details are reported in \citep{Romoli:2023C0}.

\subsection{Meteors and Nuclearites} \label{sec:Meteorsand}
To our knowledge, Mini-EUSO is the first space mission that allows a systematic study of meteors, including primarily the determination of the meteor flux in a wide range of magnitudes and the measurement of meteor light curves, as well as, possibly and in some cases at least, a computation of the original heliocentric orbits of the meteoroids.
The expected sensitivity of Mini-EUSO has been summarized 
in~\citep{ABDELLAOUI2017245} showing the capability to measure up to magnitude +5 with significant statistics (2.4 meteors/min).

\begin{figure}[h]
\centering
\includegraphics[width=0.9\columnwidth]{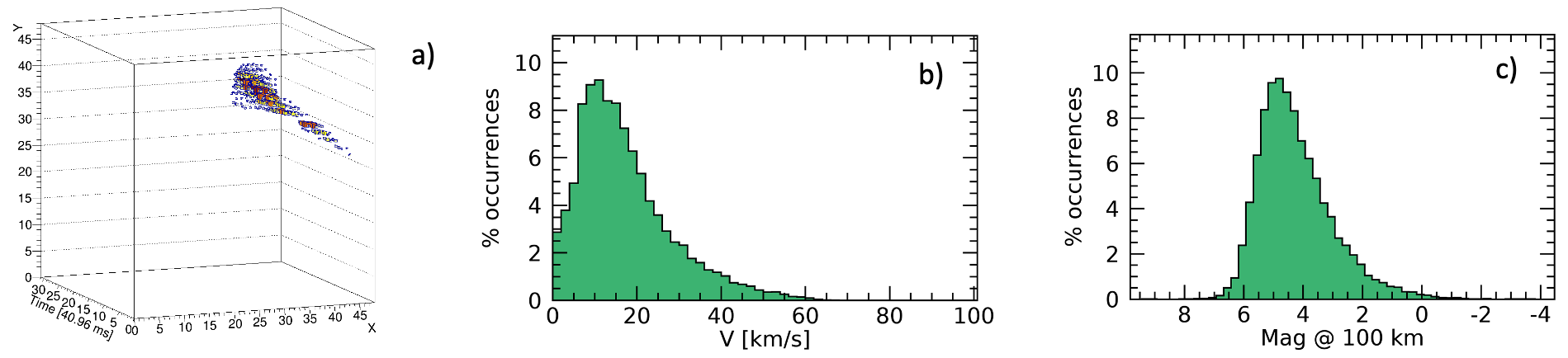}
\caption{Panel a) a meteor track as observed by Mini-EUSO at a 40.96 ms sampling rate. The X and
Y axis represent the pixels of the focal surface plotted versus time. The color scale and the size of
boxes correspond to the number of counts deposited in a pixel. Panel b) distribution of the horizontal speed
V at a 100 km reference altitude of the meteors detected by Mini-EUSO. 
Panel c) minimum absolute magnitude $M$ estimated for the events detected by Mini-EUSO. Images taken from~\citep{dario-tesi}.
}
\label{fig:meteors}       
\end{figure}

So far, in the current dataset, a total of $\sim$24,000 events were classified by Mini-EUSO as meteor events. Figure \ref{fig:meteors} shows on panel a) a meteor track as observed by Mini-EUSO at a 40.96 ms sampling rate (D3 acquisition mode). Panel b) displays the distribution of the horizontal  speed $V$ at a 100 km reference altitude of the complete sample of meteors detected by Mini-EUSO. As Mini-EUSO lacks of the stereoscopic vision, a 3-d computation of the velocity can not be performed.
Panel c) shows instead the minimum absolute magnitude $M$ distribution of the same sample.
These are just some of the results obtained by Mini-EUSO on the meteor studies. All the results and the details of the analysis can be found in~\citep{dario-tesi}, together with the results for the search of interstellar meteors.

The observing strategy developed to detect meteors may also be applied to the detection of macroscopic dark matter, generally called macros~\citep{10.1093/mnras/stv774}, which have higher velocities and a wider range of possible trajectories, but move well below the speed of light and can therefore be considered as slow events for Mini-EUSO \citep{Casolino:2023qE}.
The fact that no meteor-like event has been observed above 72 km/s speed (the maximum velocity allowed to a solar system meteor), and by taking into account the duty cycle obtained for UHECRs, it is possible to determine a preliminary estimation of the exposure for nuclearites according to the model proposed in~\citep{derujula}.
The expected limit in flux at 90\% c.l. would be 1.7$\times$10$^{-20}$cm$^{-2}$~s$^{-1}$~sr$^{-1}$ for the meteor dataset already analysed. Its extrapolation to the entire dataset under the assumption of no detection
would be 5.9$\times$10$^{-21}$cm$^{-2}$~s$^{-1}$~sr$^{-1}$ for masses above 50 gr. These values would be online with the predictions performed pre-flight on the basis of the simulations conducted in~\cite{ABDELLAOUI2017245} and, therefore, confirm experimentally those expectations. At the same time they would represent the most stringent limits so far obtained in the search for this hypothetical form of matter.



\section{Conclusions}
At the time of writing, the Mini-EUSO telescope has been taking data on board the ISS for 4 years and has proved to be an invaluable instrument to address the detection capability of a future, larger space-based detector, like K-EUSO or POEMMA. Mini-EUSO has measured the terrestrial UV background with unparalleled precision and its data have been used to estimate the effective duty cycle of a future space-based UHECR detector. A few preliminary results have been reported regarding the study of ELVES, meteors and search for nuclearites. These results confirm the multi-disciplinarity of a space-based observatory for UHECRs.

\section{Acknowledgements}
The authors acknowledge all members of the JEM-EUSO Collaboration.
This work was supported by the Italian Space Agency through the agreement n.~2020-26-Hh.0, by the French space agency CNES, and by the National Science Centre in Poland grants 2017/27/B/ST9/02162 and 2020/37/B/ST9/01821.
This research has been supported by the Interdisciplinary Scientific and Educational School of Moscow University ``Fundamental and Applied Space Research'' and by  Russian State Space Corporation Roscosmos. The article has been prepared based on research materials collected in the space experiment ``UV atmosphere''.
We thank the Altea-Lidal collaboration for providing the orbital data of the ISS.

\bibliography{my-bib-database}

\input{JEM-EUSO_Authors_July2023_final_v2.tex}



%
%
%

\end{document}

%% file: JEM-EUSO_Authors_July2023_final_v2.tex
\newpage
{\Large\bf Full Authors list: The JEM-EUSO Collaboration\\}

\begin{sloppypar}
{\small \noindent
S.~Abe$^{ff}$, 
J.H.~Adams Jr.$^{ld}$, 
D.~Allard$^{cb}$,
P.~Alldredge$^{ld}$,
R.~Aloisio$^{ep}$,
L.~Anchordoqui$^{le}$,
A.~Anzalone$^{ed,eh}$, 
E.~Arnone$^{ek,el}$,
M.~Bagheri$^{lh}$,
B.~Baret$^{cb}$
% ...
}
\end{sloppypar}
\vspace*{.3cm}

{ \footnotesize
\noindent
% ...
$^{cb}$ Universit\'e de Paris, CNRS, AstroParticule et Cosmologie, F-75013 Paris, France\\
% ...
$^{ed}$ Istituto Nazionale di Fisica Nucleare - Sezione di Catania, Italy\\
% ...
$^{eh}$ INAF - Istituto di Astrofisica Spaziale e Fisica Cosmica di Palermo, Italy\\
% ...
$^{ek}$ Istituto Nazionale di Fisica Nucleare - Sezione di Torino, Italy\\
$^{el}$ Dipartimento di Fisica, Universit\`a di Torino, Italy\\
% ...
$^{ep}$ Gran Sasso Science Institute, L'Aquila, Italy\\
% ...
$^{ff}$ Nihon University Chiyoda, Tokyo, Japan\\ 
% ...
$^{ld}$ University of Alabama in Huntsville, Huntsville, AL, USA\\
$^{le}$ Lehman College, City University of New York (CUNY), NY, USA\\
% ...
$^{lh}$ Georgia Institute of Technology, USA\\
% ...
}